\documentclass[12pt,a4paper]{article}
\usepackage{a4wide}
\usepackage{latexsym}
\usepackage{epsf}
\usepackage{amssymb}
\usepackage[all]{xy}
\usepackage{epsfig}

\def\pd{\partial}
\def\a{\alpha}
\def\b{\beta}
\def\g{\gamma}
\def\d{\delta}
\def\m{\mu}
\def\n{\nu}

\def\s{\sigma}
\def\e{\epsilon}
\def\r{\rho}

\def\RR{\mathbb{R}}
\def\CC{\mathbb{C}}
\def\ZZ{\mathbb{Z}}

\def\be{\begin{equation}}
\def\ee{\end{equation}}
\def\bea{\begin{eqnarray}}
\def\eea{\end{eqnarray}}
\def\HS{{\cal{H}}}

\def\KK{{\cal{K}}}

\begin{document}
\begin{flushright}
IFT-UAM/CSIC-04-46\\
hep-th/0410060\\
\end{flushright}

\vspace{1cm}

\begin{center}

{\bf\large Gravitational monopoles, anomalies and M$E_8$ bundles}

\vspace{.7cm}

{\bf Juan Jos\'e Manjar\'{\i}n}
\footnote{E-mail: {\tt juanjose.manjarin@uam.es}} \\

\vspace{1cm}

{\it   
 Instituto de F\'{\i}sica Te\'orica, C-XVI,
  Universidad Aut\'onoma de Madrid \\
  E-28049-Madrid, Spain}\footnote{Unidad de Investigaci\'on Asociada
  al Centro de F\'{\i}sica Miguel Catal\'an (C.S.I.C.)}

\vskip 1.8cm


{\bf Abstract}
\end{center}

In this paper we try to clarify the physical meaning of the gauge theory that underlies the K-theoretical classification of RR charges in type IIA. Our main tool are the conditions for the cancellation of the Freed-Witten global anomaly when we take into account the effects of a flat and a general B-field. In each case we will see how K-theory captures some eleven dimensional information. In the first case and studying the electric properties of the D6-brane we see an eleven dimensional $U(2)$ gauge symmetry, while the second can be related to an $E_8$ theory. Moreover, in the reduction from the general to the flat case, we find that the Romans' mass gives the number of unstable intial D9-branes.

\newpage
\section{Introduction}\label{intro}

The physical meaning of the gauge group underlying the K-theoretical classification of RR-charges and fluxes, \cite{Min.Moo, Wit.1, Hor, Moo.Wit, Fre.Hop}, is still unclear. However, as the characteristic classes appearing in K-theory are related to unitary (and orthogonal for real K-theory) groups, there is a natural link with the Chan-Paton bundles living on the world volume of the D-branes. In particular, the RR-charges of type IIA, IIB and I are classified respectively in terms of $K^{-1}(X)$, $K(X)$ and $KO(X)$.

The way K-theory enters in string theory can be seen from different points of view. The two more relevant for our purposes will be the conditions for the cancellation of global anomalies in the string partition function with D-brane boundary conditions, the Freed-Witten anomaly, \cite{Fre.Wit, Mal.Moo.Sei.1}, and the Sen-Witten description of tachyon condensation processes, \cite{Wit.1, Sen.1, Sen.2, Sen.4, Sen.5}.

In this last construction one takes into account the tachyonic instabilities that arise in the world-volume of certain bound states of D-branes or in non-BPS D-branes, so that the tachyon can condense in lower dimensional states carrying D-brane charges (see \cite{Ols.Sza, Man} and references therein).

In this paper, we will focus in type IIA string theory where, in order to obtain all the lower dimensional stable states, we need to assume the existence of an unstable ten dimensional D9 filling brane, \cite{Hor}. We also assume that this D9-brane behaves essentially as any other D-brane, i.e. it carries a $U(1)$ Chan-Paton degree of freedom and in the presence of a stack of D9-branes, this $U(1)$ is enhanced to $U(N)$. Since these D9-branes do not carry any conserved charge, we can create or annihilate any number of them from or to the vacuum, with no problem of RR tadpoles, which implies that we can have an arbitrary number of them. 

In fact, the precise definition of the characteristic classes of $K^{-1}(X)$ is given in terms of homotopy classes of maps to $U(\infty)$, the infinite unitary group, which implies an infinite number of D9-branes. This is the most general situation, and its possible physical relevance could be traced back to the second quantization in field theories.
 
However, in the construction of string solitons, we only need a finite (and definite) number of D9-branes (see section \ref{thopol}). This seems to imply that it suffices with a finite version of this group. However, there are some subtle questions regarding the discrete $\ZZ_2$ symmetry in the spectrum of the tachyon field, \cite{Wit.4, Wit.5, Man}, and the one needed to define the D8-brane charge, \cite{Hor}, that need the whole group.

One of the predictions made in this classification is that the D6-brane can be seen as a non-abelian 't Hooft-Polyakov monopole for a $U(2)$ gauge theory broken down to $U(1)\times U(1)$, \cite{Hor}. The purpose of sections \ref{mthelch} and \ref{nahhitdon} is to work out the exact characterization of this non-abelian nature of the D6-brane. This is done using the dual version of this soliton as an 11d KK-monopole, which is a purely gravitational solution of M-theory when one dimension has an $S^1$ topology, \cite{Tow.1, Tow.2, Ber.Jan.Ort, Ima}.

The main point in this construction will be the search of the possible electric degrees of freedom carried by this non-abelian soliton, \cite{Gom.Man.2, Gom.Man.1, Man}. We identify this electric charge in terms of a pure gauge deformation given by a component of the eleven dimensional 3-form, $C^{(3)}$, along the compact direction of Taub-NUT, in a construction that resembles the Julia-Zee dyon, \cite{Jul.Zee}.

One may think that including this 3-form would result into a noncommutative deformation in the lower dimensional theory. However, as we will see by a reconstruction of the moduli space and in terms of the Nahm-Hitchin-Donaldson (NHD) construction,\cite{Nah.1, Nah.2, Nah.3, Hit.2, Don}, since this field enters in the construction as the ``temporal'' component of the gauge field, we find that the gauge group for this monopole is $SU(2)$, with no central extension (as should be the noncommutative case \cite{Gom.Man.2}).

A ``spin-from-isospin'' construction allows us to see this $SU(2)$ gauge group as one of the two invariant subspaces of the $SO(4)$ group of rotations in Taub-NUT. Moreover, we can see it as the part of the $U(2)$ isometry group acting on the ${\RR}^3$ part of this space, while the extra $U(1)$ acts as motions along the compact NUT direction, i.e. in the one we can identify as the eleventh direction of space-time. Thus we can naturally relate the geometry of Taub-NUT with the geometry of the moduli space of a non-abelian monopole. Moreover, the reconstruction of the moduli space via the NHD construction allows us to identify motions along the compact NUT direction as generators of the electric charge for this gravitational monopole.

However, from the point of view of eleven dimensions, the whole gauge group is $U(2)$. This is a reflection of the fact that in our case the ``invariant direction'' in the NHD construction is physical and, following the usual KK interpretation, there is a $U(1)$ symmetry associated to it. This $U(2)$ structure should be compared with the $U(2)$ gauge group appearing in the ten dimensional K-theory discussion. In fact, as we will see in section \ref{globanom}, we can map one group to the other.

This can be understood from the interpretation of the $\pi_3$ in K-theory as the characteristic class for the Hopf bundle of the NUT direction over the sphere at infinity surrounding the D-brane, implying that the K-theoretical $U(2)$ gauge group lives in eleven dimensions. As, we will see, a careful analysis of the conditions for the cancellation of global anomalies leads to the same conclusion.

The previous analysis is done when the B-field is pure torsion, \cite{Don.Kar, Kap, Wit.1}. This is perfectly natural for the D6-brane, where this condition is dictated by geometry, however, if we want to understand the initial state, we should relax this condition and include general B-fields.

In section \ref{ntbf} we review the classification of general B-fields and how they modify the K-groups by including a twist in terms of $PU(\HS)$, the group of unitary projective operators in the (separable) Hilbert space $\HS$, \cite{Ros, Bou.Mat}. The interesting point is that, since this group is, up to dimension 14, homotopically equivalent to $\Omega E_8$, the based loop group of $E_8$, we can see the topological part of this K-theoretical classification as associated with gauge transformations of this group, \cite{Ada.Evs.1, Evs.1, Mat.Sat}.

Finally we will see how this $E_8$ formalism enters in the ten dimensional cancellation of global anomalies. In this case, we will see that the reduction of the B-field to a torsion class, which physically implies a reduction to a finite number of D-branes, can be seen as a central extension of $\Omega E_8$, $\widehat{\Omega E}_8$. It is in this context that the central charge of $\Omega E_8$ gives the number of D9-branes needed in the K-theoretical classification. 

Moreover, in the construction of the connection of this $\widehat{\Omega E}_8$ bundle, we will identify the connection associated to the $U(1)$ central extension as the $C^{(1)}$ in the Romans' formulation of type IIA massive SUGRA. This result gives further support to the conjectured relation between the central charge and Romans' mass, \cite{Ada.Evs.1, Evs.1}. 

The main difference of our analysis is that we can identify the central charge as coming from $\widehat{\Omega E}_8$, not from $\widehat{LE}_8$, the centrally extended free loop group, which has in turn a more natural relation with the dimensional reduction, see \cite{Gar.Mur, Ber.Var}.

However, a proper treatment of the problem of the anomaly cancellation should probably be done in terms of K-homology, \cite{Bro, Har.Moo, Mat.Sin, Sza}. In this paper we will present some preliminary ideas concerning this subject, leaving a more concrete realization for future work.

\section{M-Theory and Electric Charge}\label{mthelch}

In this section we will study the possible existence of an electric charge for the D6-brane. This question arises from the classification of RR-charges in terms of K-theory, where the D6-brane can be seen as a 't Hooft-Polyakov monopole.

\subsection{The D6-brane as a 't Hooft-Polyakov monopole}\label{thopol}
For type IIA, D9-branes are non-BPS and therefore are unstable. This instability is due to the existence of an open tachyon field transforming in the adjoint representation of the $U(2N)$ gauge group associated with $2N$ D9-branes. In what follows we will assume that a vev for the tachyon field is spontaneously generated inducing a breakdown of the gauge symmetry from $U(2N)$ to $U(N) \times U(N)$. Hence the corresponding vacuum manifold is

\be
\label{vmftta}
{\cal{V}}_{IIA}(2N)=\frac{U(2N)}{U(N)\times U(N)},
\ee

\noindent The topological charge is given in terms of homotopy classes of the vacuum manifold that can be easily interpreted in terms of K-theory, namely 
\be
\label{rwhg}
K^{-1}(X)=K^{-1}\left( S^{n+1}\right)=\pi_n\left({\cal{V}}_{IIA}(2N)\right)=\left\{\matrix{\ZZ, & N=2^{(n-2)/2}, \cr 0, & N=2^{(n-3)/2}.}\right.
\ee

\noindent corresponding to topologically stable solitons of space codimension $n+1$. Thus, and according to (\ref{rwhg}), the D6-brane corresponds to $N=1$ $n=2$, i.e to an initial configuration of two unstable D9-branes and space codimension three. The relevant sequence of homotopy groups for the D6-brane will be

\be
\label{hsftdsb}
\pi_3(U(2))=\pi_2\left({\cal{V}}_{IIA}(2)\right)=\pi_1(U(1))=\ZZ,
\ee

The tachyon field

\be
\label{tfftta}
T=\Gamma_mx^m,
\ee

\noindent where the $\Gamma_m$ are the usual gamma matrices, defines the map

\be
\a=-e^{i\pi T}\,:\, X_{cpt}\longrightarrow U(2N).
\ee

\noindent where, for solitons of space codimension $n+1$, we have $X_{cpt} = S^{n+1}$, so that $T$ is indeed a generator for $K^{-1}(B^{n+1},S^{n})$ \footnote{Let us briefly recall the main steps of the K-theory construction we are using. With a configuration of $N$ D9-branes we associate a $U(N)$ vector bundle $E$ and the open tachyon $T$. Following Karoubi's definition we associate solitonic D-branes of space codimension $n+1$ with elements of $K^{-1}(B^{n+1},S^{n})$ with $\a = -e^{i\pi T}$ such that $\a|_{S^{n}} =1$.}, \cite{Hor}.

Imposing the finite energy condition over the tachyon field in the $U(2)$ gauge theory over the two D9-branes,we obtain for the gauge potential

\be
\label{gpftdsb}
A_i=\frac{1}{|x|^2}\left(1-\frac{|x|}{\sinh(x)}\right)\Gamma_{ij}x^j,
\ee

\noindent where $\Gamma_{ij}=\frac{1}{2}(\s_i\s_j-\s_j\s_i)$.

The set of equations (\ref{hsftdsb}), (\ref{tfftta}) and (\ref{gpftdsb}) define the D6-brane as a 't Hooft-Polyakov monopole, i.e as a magnetically charged soliton of space codimension three, for a ten dimensional $U(2)$ gauge theory spontaneously broken to $U(1)\times U(1)$.

\subsection{The D6-brane as a KK monopole}\label{kkmono}

Opposite to this non-abelian description, we have the usual supergravity one, where the D6-brane can be seen as a KK-monopole in eleven dimensions. On general grounds, the KK-monopole in $d$ dimensions is metrically described by

\be
\label{msftkm}
ds_d^2=ds^2_{d-4}+ds^2_{TN},
\ee

\noindent where $ds^2_{d-4}$ represents the core of the monopole and $ds^2_{TN}$ is the transverse space, given by a Taub-NUT metric, i.e. it has an isometric direction. 

The usual KK reduction mechanism states that in $d-1$ dimensions we see the component of the metric along the isometric direction as a $U(1)$ vector field. In this case it reads

\be
\label{tngp}
A_\m=g_{\m 4}^{(TN)}=4m\left( 1-\cos\theta\right)d\phi,
\ee

\noindent where $m$ is the NUT charge. From (\ref{tngp}) one can compute the magnetic field and charge associated to this monopole. 

As a prediction of S-duality, the spectrum of the theory must contain states with both, magnetic and winding charge, \cite{Sen.3, Gre.Har.Moo}. The origin of this winding charge can be traced back to the H-monopole and, therefore, to the presence of the Kalb-Ramond B-field. As the transverse space of (\ref{msftkm}) is four dimensional one could expect to have four translation zero modes, however, the isometry along $x^4$ implies that we have only three zero modes. The extra zero mode comes from the massless spectrum of the theory and the only harmonic 2-form field in Taub-NUT space

\be
\label{ohzftn}
B=\frac{c}{4m}d\left[\frac{r}{r+4m}\left( dx^4+4m(1-\cos\theta)d\phi\right)\right]=cd\chi,
\ee

\noindent which is seen as a pure gauge non-vanishing at infinity. This deformation is reminiscent to that needed to obtain the electric charge in the Julia-Zee dyon

As a consequence, we get as the moduli of the KK-monopole $\RR^{3} \times S^{1}$, namely the same moduli that for the 't Hooft Polyakov monopole. Notice that, in this case, we have used, in order to get this moduli, two ingredients, the $U(1)$ gauge connection obtained by dimensional reduction and the Kalb Ramond $B$ field. 

This analysis made in ten dimensions can be directly generalized to eleven dimensions and, therefore, to the D6-brane. However, in eleven dimensions we do not have a 2-form and therefore any form of {\em electric charge} of the D6-brane should be associated with the eleven dimensional 3-form, $C^{(3)}$, \cite{Gom.Man.2, Gom.Man.1}. In analogy with the ten dimensional case, we will refer to the state carrying the charge of the eleven dimensional 3-form as the C-monopole.

In fact in the presence of a KK-monopole, we can decompose the 3-form, \cite{Ima} as 

\be
\label{C}
C^{(3)}_{MNP}\stackrel{KKM}{\longrightarrow}\left\{\matrix{C_{\m\n\rho} & & \cr C_{\m\n i} & = & C_{\m\n}^{(2)}A_i \cr C_{\m ij} & = & V_\m B_{ij}^{(2)} \cr C_{ijk}}\right. ,
\ee

\noindent where $x^\m=(r,\theta,\phi,x^4)$ are the Taub-NUT coordinates (with $x^4$ the eleventh dimensional coordinate) and $y^i$, $i=0,...,6$ the coordinates on the world-volume of the KK-monopole. In order to match the actions of the eleven dimensional KK-monopole and the D6-brane, one obtains that both 2-forms, $B^{(2)}$ and $C^{(2)}$, in (\ref{C}), are flat and given by

\bea
\label{tfvof}
dA\propto B^{(2)},\qquad dV\propto C^{(2)}.
\eea

As $C^{(2)}$ lives in the Taub-NUT space, we can use this two form in exactly the same way we use the Kalb Ramond field in the ten dimensional case. Therefore, $V$ will give its trivialization in terms of the vierbein of Taub-NUT. We see, on the other hand, that the main difference with the ten dimensional case is that including the 3-form gives information on the field content in the world-volume of the D6-brane.

In \cite{Ima} it was proposed a slight modification of (\ref{ohzftn}) by including a gauge freedom

\be
\label{edhf}
C^{(2)}=\frac{c}{4m}d\left[ f_1(r)dr+\frac{r}{r+4m}\left( dx^4+4m(1-\cos\theta)d\phi\right)\right].
\ee

Using (\ref{edhf}) we can identify the electric degrees of freedom with the component $C^{(2)}_{\m 4}$ of (\ref{edhf}). This gives rise to the pure gauge field

\be
\label{gfftf}
A_{\m}^{(2)}=C^{(2)}_{\m 4}=\frac{c}{(r+4m)^2}dr,
\ee

\noindent where the constant $c$ should be fixed by physical considerations. If we consider the coupling to the eleven dimensional membrane and perform the dimensional reduction, then, choosing a trajectory for which $\theta$ and $\phi$ are constant we can set

\be
c=\frac{3}{(4\pi)^2g_{M2}\a'} 
\ee

\noindent where we have introduced a $1/2\pi\a'$ factor since it is the proportionality constant in (\ref{tfvof}). This factor reduces to $c=4m$ once we pass from eleven to ten dimensional variables.

The Hodge dual of (\ref{gfftf}) is proportional to the volume form of Taub-NUT and, therefore, integration over the sphere at infinity gives rise to the electric charge.

Notice that the existence of some form of electric charge for the D6-brane, interpreted as a KK monopole in eleven dimensions, is {\em necessary} if we want to identify the D6-brane with the codimension three topologically stable soliton appearing as a consequence of tachyon condensation in type IIA string theory. In fact only in this case the moduli of both candidates for the D6-brane coincide and are equal to $\RR^{3} \times S^{1}$. We will postpone the discussion of this moduli space to next section.

\subsection{Spin from Isospin Construction}\label{spiso}

Firstly, following \cite{Hul}, we know that the magnetic charge of the KK-monopole can be computed from the totally antisymmetric part of the spin connection minus the contribution of the base space, which is given by

\be
\label{scftn}
\omega=F\wedge k,
\ee

\noindent where $k$ denotes the Killing vector along the isometric direction and $F$ is the field strength of (\ref{tngp}).

This spin connection has $SO(4)$ as group structure, the group of rotations in the Taub-NUT space-time. However, the isometry group of this space is isomorphic to $U(2)=SU(2)\times U(1)/\ZZ_2$, which we can take formally as our gauge group, therefore, let us consider the embedding of the spin connection into the $SU(2)$ subgroup, linking the isospin to one of the $SU(2)$ invariants subspaces of $SO(4)$. This can be done in terms of the 't Hooft's symbols

\be
\label{tstm}
\bar\eta^i{}_{ab}=\left\{\matrix{\bar\eta^i{}_{ab} & = & \epsilon^i{}_{ab} & (i,a,b=1,2,3) \cr \bar\eta^i{}_{4b} & = & -\delta^i{}_b & (i,b=1,2,3) \cr \bar\eta^i{}_{a4} & = & \delta^i{}_a & (i,a=1,2,3) \cr \bar\eta^i{}_{44} & = & 0 &  }\right.
\ee

\noindent The resulting $SU(2)$ gauge connection is

\be
\label{esc}
{\cal{A}}_{\m}^i=\frac{1}{2}\bar\eta^i{}_{ab}\omega_\m{}^{ab},
\ee

\noindent where $i\in U(2)$, $a\in SO(4)$ and $\m\in$ Taub-NUT.

The connection (\ref{esc}) inherits an important property of Taub-NUT and it is its (anti) self-duality in eleven dimensions. In field theory we can describe a 't Hooft-Polyakov monopole in the Prasad-Sommerfield (PS) limit in terms of an (anti) self-dual gauge connection invariant under translations in the time direction. In the gravitational case we will formally interpret this ``time invariance'' as the dimensional reduction on the isometry direction. 

In the remainder of this section, we will use the following convention of indices. The curved indices are $M=\left\{ y,\m\right\}$, where $y$ is the eleventh dimension, while the flat indices are $A=\left\{\underline y,a\right\}$. Therefore gauge field (\ref{esc}) is

\be
{\cal{A}}_M^i=\frac{1}{2}\bar\eta^i{}_{AB}e^A{}_Re^B{}_Ng^{RP}g^{NQ}\omega_{MPQ},\\
\ee

\noindent where $e^A{}_M=\left\{e^{\underline y}{}_M,\hat e^a{}_\m\right\}$ is the vierbein of Taub-NUT, that involves the eleventh dimension, and satisfies

\bea
\label{votn}
e^A{}_Me^B{}_N\eta_{AB}=g_{MN},\qquad \hat e^a{}_\m\hat e^b{}_\n\eta_{ab}=V^{-1}g_{\m\n}=\frac{r+4m}{r}g_{\m\n}.
\eea

Using (\ref{tstm}) and (\ref{votn}) we can write down the components of the gauge connection as

\bea
\label{cecg}
{\cal{A}}^i{}_y & = & \frac{4m}{(r+4m)^2}dr, \\
\label{cmcg}
{\cal{A}}^i{}_\m & = & \left( 0,\d^i{}_3,\d^i{}_2\sin\theta\right)\frac{4m\left( K(r-4m)-1\right)}{r},
\eea

\noindent where

\be
K(r-4m)=\left(\frac{4m\left( 1-\cos\theta\right)}{r\sin\theta}\right)^2+2
\ee

Comparing (\ref{cecg}) and (\ref{gfftf}) we see that the electric part of the gauge field is a pure gauge field which incorporates the data of the eleven dimensional 3-form once we {\em identify} the $SU(2)$ index with a space-time index.

In summary we observe that using a spin from isospin map we can, in pure eleven dimensional supergravity, associate the D6-brane with a non-abelian anti self-dual gauge connection with a well defined electric charge that coincides with the one derived from the 3-form C. In next section we will work out the Nahm-Hitchin-Donaldson (NHD) construction for this anti self dual gauge connection.

\section{NHD Construction and the D6-Brane Moduli}\label{nahhitdon}

Naively, the {\em moduli} of one D-brane is isomorphic to the transversal target space. This statement becomes a bit more subtle for the D6-brane. In fact if, according with the tachyon condensation approach, we interpret the D6-brane as a 't Hooft-Polyakov monopole, we should expect as the moduli not the naive transversal space $\RR^{3}$ but instead $\RR^{3} \times S^{1}$. The main goal of our previous discussion on the electric charge of the D6-brane was to provide indirect evidence on this form of the moduli. In fact the $S^{1}$ fibration of the moduli is directly associated with the existence of an electric charge.

Let us start defining the $SU(2)$ bundle on which the gauge connection (\ref{esc}) takes values. Let $H$ be the Hopf bundle over $S^2$. For a charge $k$ monopole, the direct sum $H^k\oplus H^{-k}$ defines another bundle over $S^2$ which can be extended radially over $\RR^3-\{ 0\}=S^2\times\RR^+$, \cite{Ati.Hit}. This construction gives an $SU(2)$ bundle with transition functions

\be
k_{\a\b}:U_\a\cap U_\b\longrightarrow SU(2),
\ee

\noindent where

\be
\label{fts}
k_{\a\b}=\pmatrix{ g_{\a\b} & \cr
                            & h_{\a\b}},
\ee

\noindent and we can consider

\be
\left( g_{\a\b}, h_{\a\b}\right):U_\a\cap U_\b \longrightarrow U(1),
\ee

\noindent as the transition functions for the (magnetic, electric) bundles.

It is known that the Taub-NUT metric has a coordinate singularity which can be removed by choosing two different patches and the gauge transformation 

\be
\label{gtitn}
y\rightarrow y-8m\phi,
\ee

\noindent in the intersection. This is the transformation we will choose to define the transition functions in (\ref{fts}). 

The magnetic field is defined in terms of the non-diagonal terms of the Taub-NUT metric from

\be
ds^2_{nd}=\frac{r}{r+4m}\left[ dy+4m\left( 1-\cos\theta\right)d\phi\right]^2.
\ee

The transformation (\ref{gtitn}) implies that the gauge field varies as $A_\m^{(1)}\rightarrow A_\m^{(1)}-8md\phi$ and, therefore

\be
\label{elb}
i\log h_{\a\b}=d(\d A_\m^{(1)})=-8m\phi.
\ee

The transition function $g_{\a\b}$ is a bit more subtle, since it involves a flat 2-form which, in the mathematical language, defines a flat gerbe. In general, a gerbe is defined in terms of a triple $(C_\a,\Lambda_{\a\b},g_{\a\b\g})$ where, as usual, the indices denote that they are defined on one open set and in the double and triple intersection of open sets. However, when the 2-form $C\propto dV$ is flat, the structure gets some modifications and $V$ gives a trivialization of $C$. This is translated into

\bea
\left(\d C^{(2)}\right)_{\a\b} & = & -d\Lambda_{\a\b},\\
\left(\d V\right)_{\a\b} & = & \Lambda + d\rho_{\a\b}.
\eea

\noindent and we do not care of $g_{\a\b\g}$ since it is not important for our purposes.

Using (\ref{gtitn}) we can write

\bea
\Lambda=2c\frac{4m}{(r+4m)^2}\phi dr,\qquad \rho=-2c\frac{r}{r+4m}\phi.
\eea

\noindent From gerbe theory, (see for example \cite{Hit}) it is known that the difference of two trivializations is a complex line bundle defined on a double overlap, which can be used to define the gerbe without using a 2-form. This line bundle has a connection which can be seen to be the previous $\Lambda$. In order to define the electric transition function we will take this complex line bundle and as transition function the one that defines it. This, in turn, implies that in the $r\rightarrow\infty$ limit

\be
\label{mlb}
i\log g_{\a\b}=-2c\phi.
\ee

With this we can write the transition function for the $SU(2)$ bundle on which the embedded connection (\ref{esc}) takes values, namely

\be
k_{\a\b}=\pmatrix{ e^{8m\phi i} & \cr & e^{-8m\phi i} }.
\ee

\noindent where $0\leq\phi\leq 2\pi$.

Now that we have properly identified the moduli space of the D6-brane as that for an $SU(2)$ monopole, we can extend the NHD construction to the anti self-dual gauge connection (\ref{esc}), deriving a set of monopole Nahm equations by direct dimensional reduction on the isometric direction of Taub-NUT.

The Nahm's transformation can be seen as a sort of Fourier transformation for an anti self-dual connection, where invariance in direct space is translated into dependence in dual space and viceversa. In the PS limit we take the monopole as a gauge connection invariant under time translations which can be obtained as a solution to the Bogomolnyi equations. In the Nahm-Hitchin's construction, we impose that the Fourier transformed connection only depends on the dual coordinate to time, \cite{Nah.1, Nah.2, Nah.3, Hit.2}. Moreover, Donaldson's construction allows us to see the Nahm's equations as the dimensionally reduced anti self-duality equations in Fourier space, \cite{Don}.

Our case is rather similar, the only difference being that now we work with a gravitational set up, i.e. our higher dimensional object is a gravitational instanton, the KK-monopole, and after dimensional reduction we obtain, what may be called, a gravitational monopole, the D6-brane. This means that the dimensional reduction is physical and, therefore, as our compact direction is a circle, we will get a $U(1)$ gauge symmetry.

The field strength of (\ref{esc}) can be written as

\be
F_{MN}{}^i=\frac{1}{2}\overline\eta^i{}_{AB}\pd_{[M}\omega_{N]}{}^{AB}+\frac{\epsilon^i{}_{jk}}{4}\overline\eta^j{}_{AB}\omega_{[M}{}^{AB}\overline\eta^k{}_{|CD|}\omega_{N]}{}^{CD},
\ee

We can now consider the gauge field as that associated to $\omega$ and the $C^{(2)}$ part as a deformation. This implies that we can write the Fourier transformed connection as

\be
A(s)\equiv\hat{\cal{A}}(s)^i_M\s_i\,dp^M=T^i\s_i
\ee

\noindent and, in terms of these matrices $T$ write the Nahm's equations as

\be
\frac{dT_\m}{ds}+\left[ T_y,T_\m\right]\mp\frac{1}{2}\epsilon_{\m\n\g}\left[ T^\n,T^\g\right]=0,
\ee

A very important question in the Nahm's construction is that of the boundary conditions for the $s$ coordinate. In fact the Callias' index theorem implies that $s\in[0,2]$ and the $T$ functions are analytic for $s\in(0,2)$ and have simple poles at the end points, where they can be written as

\be
\label{leftm}
T_\m=\frac{t_\m}{s}+d_\m,
\ee

\noindent and these residues $t_\m$ furnish an irreducible representation of $SU(2)$. Following \cite{Don} we can take the following combinations

\bea
\label{ronm}
\a=\frac{1}{2}\left( T_y+iT_1\right),\qquad \b=\frac{1}{2}\left( T_2+iT_3\right).
\eea

\noindent and, denoting by $a$ and $b$ the residues of $\a$ and $\b$ respectively, we can compute the eigenvalues equation for $a$ from

\be
av=-\frac{k-1}{4}v.
\ee

Now we are in position to define a Nahm complex as the triple $(\a,\b,v)$, \cite{Don, Ati.Hit}. The set of equivalence classes of Nahm complexes is naturally identified with the circle bundle ${\cal{M}}_k\times S^1$ over the moduli space of monopoles ${\cal{M}}_k$. This extra $S^1$ piece of the moduli space is the interesting part for us, since motions along it are the generators of the electric charge of the monopole.

Two Nahm complexes are equivalent if there is a continuous map $g\,:\,[0,2]\rightarrow GL(k,\CC)$ smooth in the interior and such that $i$) $g(\a,\b)=(\a',\b')$ in $(0,2)$, where

\bea
\a' & = & g\a g^{-1}-\frac{1}{2}\frac{dg}{ds}g^{-1} \\
\b' & = & g\b g^{-1}
\eea

\noindent $ii$) $g(2-s)=g^T(s)^{-1}$ and $iii$) $g(0)v=v'$. This implies that, writing $g(s)=e^{i\phi(s-1)}$, we obtain that a change in $v'=e^{-i\phi}v$ results in a change $\a'=\a-\frac{i}{2}\phi$, which explains the $S^1$ bundle structure.

The matrices (\ref{ronm}) allows, on the other hand, to identify the hyperk\"ahler structure of the moduli space by writing the Nahm's equations in terms of them and interpreting the resulting equations as the moment maps for this structure. A local solution of the complex equation can be written as

\bea
\label{lstne}
\a=\frac{1}{2}g^{-1}\frac{dg}{ds},\qquad \b=g^{-1}\b'g
\eea

\noindent which allow us to identify $\a$ as a locally pure gauge parameter.

From Donaldson's construction, we see that the $S^1$ moduli is generated in terms the eigenvector, $v$, of the residues of $\a$ at the boundary of the definition interval of the matrices $T_i$.

In the gravitational construction, since from (\ref{cmcg}) we see that the ${\cal{A}}_1^i$ component of the gauge field vanishes, this $\a$ parameter is related to the pure gauge component of the eleven dimensional 3-form, $C^{(3)}$, along the compact direction (\ref{gfftf}),
 
\be
\a=\frac{1}{2}\hat{\cal{A}}_y.
\ee 

The fact that this component of the gauge field is a pure gauge, as is also the case in the usual monopole, is in accordance with (\ref{lstne}). 

Therefore, we can conclude that the $S^1$ bundle of the moduli space of the D6-brane is directly associated with the eleven dimensional 3-form, which gives rise to its ``electric'' charge. Moreover, this bundle (and so the electric charge) is a reflection of the existence of an extra compact dimension, which is naturally associated with the Taub-NUT direction, i.e. with the eleventh dimension.

This can be seen from the Donaldson's theorem relating the monopole moduli space with set space of rational functions. The scattering function for this KK-monopole can be written as

\be
f(z)=\frac{Ae^{-\hat{\cal{A}}(1)_y}}{z-z_0},
\ee

\noindent where $A$ is an amplitude, that describes a monopole located at $\left(-\frac{1}{2}\log\left| Ae^{-\hat{\cal{A}}(1)_y}\right|,z_0\right)$ with phase $Arg\left( Ae^{-\hat{\cal{A}}(1)_y}\right)$, so that a change in $S^1$, $v\rightarrow e^{i\phi}v$, results in a change $Ae^{-\hat{\cal{A}}(1)_y}\rightarrow Ae^{-\hat{\cal{A}}(1)_y+i\phi}$, generating the electric charge of this monopole.

Therefore, we see the electric charge as translations in the Fourier transformed component of the "time" component of the gauge field. In direct space this corresponds to the gauge transformations of ${\cal{A}}_y$, that is a pure gauge, which, taking into account the reconstruction of the moduli space that we have carried at the begginning of this section, can be seen as motions (\ref{gtitn}) along the eleventh dimension.

It should be noted that this charge is not the electric charge in the field theory sense. It is an electric charge by construction, i.e. the charge that appears in the theory of a magnetic monopole if one includes the effect of a pure gauge field along a privileged direction, which is the one along which the monopole is translationally invariant. However, in our case this a spatial direction and not the time coordinate, which remains in the core of the monopole.

\section{The global anomaly and the electric charge}\label{globanom}

In the previous sections, we have carried out a SUGRA analysis of the moduli space and the charges of the D6-brane, and we have found that it can be seen as a non-abelian 't Hooft-Polyakov monopole, with an electric charge associated with motions along the eleventh compact dimension. Now we want to make contact with K-theory.

The first important point is the one concerning the gauge groups. From the SUGRA construction we have seen that the gauge group underlying the non-abelian KK-monopole is $SU(2)$. However, K-theory predicts a $U(2)$. Therefore, we may wonder to what extent both constructions are the same.

Indeed it is not difficult to see that we can set an application that takes the SUGRA to the K-theoretic construction. On the K-theoretical classification the tachyon is a Higgs' like excitation and so, for a stable vortex soliton of codimension $m+1$, it defines a map from the sphere at infinity surrounding the soliton into the vacuum manifold, $S^m\rightarrow{\cal{V}}_{IIA}$, with a characteristic class in $\pi_m({\cal{V}}_{IIA})$.  The tachyon can be written as

\be
T(x)=f(|x|)\Gamma_i x^i,
\ee

\noindent where the $\Gamma_i$ are the Dirac matrices of $SO(m+1)$, $x^i$ are the coordinates in the transverse space to the soliton and $f(|x|)$ is a convergence factor. This homotopy group can be mapped into $\pi_{m+1}(U(2N))$, that defines the state prior to the breaking of the symmetry. In order to compute this classes we need a map

\be
\label{mfkt}
{\cal{U}}\,:\, S^{m+1}\longrightarrow U(N).
\ee

Heuristically, as the tachyon field is an adjoint valued field, we can think of this $U(N)$ as the exponential of this field, so we can write

\be
\label{tfmfkt}
{\cal{U}}=-e^{i\pi T}.
\ee  

The subtlety comes from the sphere, however, using $T^2=|x|^2$, we can see, \cite{Hor, Dub.Fom.Nov}, that (\ref{tfmfkt}) corresponds to the map (\ref{mfkt}).

In the case of the D6-brane, the group of rotations is $SO(3)$ which has as non-trivial double covering group $Spin(3)$, which is, in turn, isomorphic to the group of unit quaternions $\mathbb{H}$, i.e. to $SU(2)$. This $SU(2)$ is the gauge group of the spin connection once we project it in terms of the "spin-from-isospin" construction and it is also the spinor representation that we identify, via the construction just explained, with the $U(2)$ Chan-Paton gauge group. Therefore, we can interpret the exponential map (\ref{mfkt}) as the map from SUGRA to K-theory.

The essential point in this map to K-theory is that (\ref{mfkt}) has a characteristic class in $\pi_3(U(2))$, and we can read the sequence (\ref{hsftdsb}). There we find a $\pi_1$ that is given by the large gauge transformations in the equator of the $S^2$ surrounding the D6-brane at infinity, and $\pi_2$ that measures the triviality of the gauge bundle. 

\begin{figure}[h]
\begin{center}
\leavevmode
\epsfxsize=8cm
\epsffile{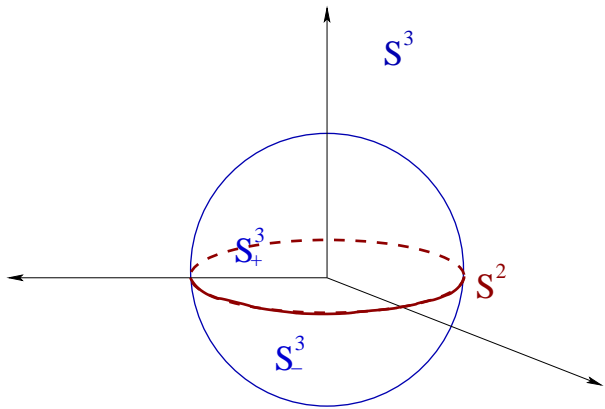}
\caption{\it Gauge transformations for the D6-brane}
\label{f1}
\end{center}
\end{figure}

K-theory lies in the $\pi_3$, which is usually associated with the compact support of the construction. However, for the special geometry of the case at hand, we can interpret it as giving information of the $S^1$ fiber of the eleventh dimension. To see it, let us remember that at infinity, where we recover the $SU(2)$ gauge group, Taub-NUT can be seen as a Hopf fibration

\be
S^3\longrightarrow S^2,
\ee

\noindent with the Taub-NUT $S^1$ as fiber (see figure \ref{f1}). 

Therefore, interpreting this $S^3$ as the one in (\ref{hsftdsb}), we are led to interpret the $U(2)$ gauge group on the world-volume of the two D9-branes as the $U(2)=SU(2)\times_{\ZZ_2}U(1)$ group of isometries. This interpretation does not support the consideration of the $S^1$ extension needed to define the $K^{-1}(X)$ group as the eleventh dimension. In fact, as shown in appendix \ref{k1tok}, this extension is needed to define an $S^4$ manifold that serves as classifying space.

However, it should also be stressed that this interpretation holds when we include the effects of the B-field. Namely, as we have seen in the previous sections, it is the eleven dimensional 3-form the one that enters as motions along the compact direction and, therefore, only when we include it, and we have to remember that it is a pure gauge field, we see the effects of the $S^1$ fibration. In order to give support to this lift of the Chan-Paton gauge group to eleven dimensions, we can study the conditions for the cancellation of global anomalies in the presence of a torsion B-field. 

It is known that including the B-field can turn the gauge theory in the world-volume of the D-branes into a noncommutative one. This can be understood, from a group theoretical point of view, in terms of Azumaya algebras, \cite{Kap, Man}. In the end, this algebraic structure implies that, in the case of a torsion B-field, the gauge theory for a system of $N$ D-branes is no longer a $U(N)$ one, but rather $U(N)/U(1)=SU(N)/\ZZ_N$.

However, we have seen in the previous sections that, in our construction, the B-field does not produce a noncommutative deformation, so that the $U(2)$ does have an $SU(2)$ structure. The condition for this structure can be deduced from the cancellation of global anomalies. Let us consider the world-volume theory on the two D9-branes with a flat B-field present. In \cite{Kap} it was demonstrated that the non-abelian condition for the cancellation of global anomalies is

\be
\label{cnaga}
\b(y)+W_3=[H],
\ee

\noindent where $W_3\in H^3(X,\ZZ)$ is the third Stiefel-Whitney class, $[H]$ is the characteristic class of the B-field and $\b(y)$ is the Bockstein homomorphism of $y\in H^2(X,\ZZ_2)$ (that represents the 't Hooft flux) in terms of the long exact sequence

\be
\dots\rightarrow H^2(X,\ZZ)\rightarrow H^2(X,2\ZZ)\rightarrow H^2(X,\ZZ_2)\stackrel{\b}{\rightarrow} H^3(X,\ZZ)\rightarrow\dots
\ee

The gauge structure of the world-volume theory of the two D9-branes is such that it is a $U(2)$ gauge theory without vector structure. The condition, \cite{Kap}, to have this $SU(2)$ structure is

\be
\b(y)=0,
\ee

\noindent which is only possible if $y=c$ mod 2, where $c\in H^2(X,\ZZ)$, i.e. it is the reduction mod 2 of the first Chern class of a complex line bundle associated. 

These conditions define a $Spin^c$ structure and moreover, we see that, as should be, the B-field does not produce a noncommutative deformation, since this would be reflected in the form of a central extension in the algebra (see \cite{Gom.Man.2}).

This $Spin^c$ structure can be understood as arising from $U(N)$ as follows (see \cite{Ati.Bot.Sha}). Since the $U(N)$ gauge group defines a principal bundle of unitary frames

\be
U(N)\stackrel{\a}{\longrightarrow} F^u\longrightarrow X,
\ee

\noindent where $X$ is the base manifold, we can define canonically a $Spin^c(2N)$ structure in terms of the homomorphism

\be
l\,:\,U(N)\longrightarrow Spin^c(2N),
\ee

\noindent that fits in the commutative diagram

\be
\xymatrix{
Spin^c(2N) \ar[r]^{\rho\hspace{0.4cm}} & SO(2N)\times U(1) \\ 
U(N) \ar[u]_l \ar[ur]  &
},
\ee

\noindent where the $\rho$ defines the $Spin^c$ structure in terms of the short exact sequence

\be
0\rightarrow \ZZ_2\rightarrow Spin^c(2N)\rightarrow SO(2N)\times U(1)\rightarrow 1.
\ee

From these considerations we can write the following sequence

\be
\label{som}
U(2)_{CP}\stackrel{l}{\longrightarrow}Spin^c(4)=SO(4)\times_{\ZZ_2} U(1)\stackrel{\s}{\longrightarrow}SU(2)\times_{\ZZ_2} U(1)=U(2)_{Isom},
\ee

\noindent where $U(2)_{CP}$ is the Chan-Paton gauge group, $\s$ is the ``spin-from-isospin'' projection and the $U(2)_{Isom}$ is the isometry group. Therefore, the trace of an eleventh dimension can be seen in ten dimensions as the existence of this $Spin^c$ structure, whose $SO(4)$ part is described by the spin connection (\ref{scftn}) and the $U(1)$ factor comes from the complex structure and can be associated with the killing isometry, by means of the ``spin from isospin'' construction of section \ref{spiso}.

Therefore, the main conclusion we obtain in this section is that the conditions for the cancellation of global anomalies in the string world-sheet path integral can be seen as the conditions to lift to eleven dimensions. Moreover, in the special case of a Taub-NUT geometry, the extra $U(1)$ that allows us to define the $Spin^c$ structure can be associated with the residual gauge symmetry of the compact eleventh dimension and, so, with the electric charge of the D6-brane.

\section{Non-torsion B-fields and $E_8$ bundles}\label{ntbf}

In the previous section we assumed that the B-field was flat. This is indeed the natural condition for D6-brane/KK-monopole, \cite{Gom.Man.1}, where the explicit form of this field is dictated by the geometry of space-time and by the condition $G^{(4)}=0$ in 11d. 

However, the initial state in the K-theoretical construction is a stack of unstable D9-branes. In this case we cannot impose any a priori condition on the fields and, therefore we should consider the most general situation and take the B-field living in the world-volume of these D-branes as non-torsion.

We have also seen that including the B-field leads to uncover a gauge structure in eleven dimensions, even when its cohomology class vanishes and we can consider the ordinary K-theoretical classification. In this section we will see that, topologically, the gauge structure associated with a non-torsion B-field can be related to previous constructions based on the existence of an eleven dimensional $E_8$-bundle, see for example \cite{Dia.Moo.Wit.1, Ada.Evs.1, Evs.1, Mat.Sat}.

As a 2-form, the B-field has a class $[H]\in H^3(X,\ZZ)$, where $H^3(X,\ZZ)$ is the third \v Cech cohomology group in sheaf cohomology. Following a theorem of Dixmier-Douady (DD) for continuous-trace algebras, \cite{Dix.Dou}, we can see this class as defining a principal bundle with fiber the projective unitary group $PU(\HS)=U(\HS)/U(1)$

\be
\label{pbd}
PU(\HS)\rightarrow P_{[H]}\stackrel{\pi}{\rightarrow} X.
\ee

In the presence of this cohomology class we can define K-theory (see \cite{Ros, Bou.Mat, Man})

\be
\label{tkg}
K^j\left(X,[H]\right)=K_j\left(\Gamma_0\left(X,{\cal{A}}_{[H]}\right)\right),
\ee

\noindent in terms of the algebra of sections vanishing at infinity, $A_{[H]}=\Gamma_0\left( X,P_{[H]}\times_{PU(\HS)}\KK\right)$, of the associated vector bundle, where $\KK$ is the set of compact operators and $PU(\HS)$ acts by $^*$-automorphisms given by the adjoint map, $Ad\,:\,T\longrightarrow gTg^{-1}$.  In particular, Rosenberg showed, \cite{Ros}, that

\be
\label{tkgh}
K^{-1}\left(X,[H]\right)=\left[ P_{[H]},U_{cpt}\right]^{PU},
\ee

\noindent where $U_{cpt}=\left\{ u\in U({\cal{H}})\big| u-1\in{\cal{K}}\right\}$ is the group of unitary operators in the unitalization of ${\cal{K}}$. 

The physical picture that arises from these considerations (and those made in the introduction) is that the general initial state for K-theory requires an infinite number of initial unstable D9-branes\footnote{As shown in \cite{Wit.4}, this infinite number of D9-branes is also neccesaryy so that the tachyon is homotopic to a constant at infinity.}. Of course, not all of these D9-branes carry a lower dimensional charge and they can decay to the vacuum in a process of tachyon condensation, while some others will give rise to the different stable states in Type IIA. 

The inclusion of an $H$-field is reflected in the $PU(\HS)$ twist that we have to perform, so that the new gauge group\footnote{It is worth to mention that this is the picture proposed in \cite{Har} for the gauge group of a non-commutative gauge theory in the presence of a NS5-brane charge.} in the world volume of the D9-branes is

\be
\label{vmwntbf}
{\cal{G}}=\frac{U_{cpt}}{PU(\HS)}
\ee

\noindent so that the topological charges are classified\footnote{This is why, in an abuse of notation, in \cite{Man}, (\ref{vmwntbf}) was referred to as the vacuum manifold for type IIA.} by the homotopy classes of $PU(\HS)$. In fact, any model for $K(\ZZ,2)$ works.

It is therefore tempting to conjecture, \cite{Har.Moo}, that the system can now also condense into a state carrying $H$ charge, so that we can have NS5-brane backgrounds. However, a main problem with this proposal is that, as remarked in \cite{Har.Moo}, it is difficult to make sense of the partition function, since the operators involved are not necessarily of trace class. 

In \cite{Car.Mic} it was proposed a refinement by taking into account the group $U_{res}$, defined as the set of unitary operators such that their non-diagonal blocks with respect to certain polarization $\e$ are Hilbert-Schmidt, by means of a modification of the Atiyah-Singer (AS) construction that considers the group of unitary operators that differ from the identity in a trace class operator, $U_{res}$. However, since $U_{cpt} $ and $U_{res}$ have the same homotopy \cite{Pal}, this modification does not alter the topological properties we are dealing with, thus we will insist on $U_{cpt}$ in the remainder of this paper, leaving for a future work the role of $U_{res}$.

\subsection{Tachyon condensation}\label{taquion}

We have just described the "should-be" situation before the tachyon condensation. Let us now see how this would take place. 

The tachyon, as a scalar field can be taken as a section of the bundle described in the previous section, therefore, it will be a compact operator transforming in the adjoint of $U(\HS)$. Of course, since we are in a non-commutative theory, it also satisfies a projector like equation, $T*T=T$, where ``*'' denotes the usual star product in non-commutative gauge theories, \cite{Gop.Min.Str, Das.Muk.Raj}, which, by means of the Weyl map is the usual product of operators. Therefore, the general form of the tachyon field is

\be
T=\sum_{i=1}^Na_iP^i
\ee

\noindent where $\{ P^i\}$ are mutually orthogonal projection operators and the coefficients $\{ a_i\}$ take values in the set of extremal points of the scalar potential.

The ranges of this kind of projection operators define finite dimensional bundle like objects with finite dimensional fibres. These are the so-called ``gauge bundles'', that were used in \cite{Bou.Mat} to define $K(X,[H])$ in terms of Murray-von Neumann equivalence classes. What we need is a similar definition for $K^{-1}(X,[H])$. 

We can assume that, in the presence of a non-vanishing H-field, the gauge structure in the world-volume of the D-branes is that of a gauge bundle. Therefore, following Karoubi \cite{Kar}, we can define the characteristic classes in $K^{-1}(X,[H])$ as defined in terms of pairs $({\cal{G}},\Phi)$, where ${\cal{G}}$ is a set of gauge bundles and $\Phi$ denotes the automorphisms of ${\cal{G}}$.

In order to construct these automorphisms, which we denote by $\Phi$, let us take the sections $A_{[H]}$ of the $P_{[H]}$ bundle, which are a set of functions $f_\a\,:\,U_\a\to\KK$, such that on double overlaps, satisfy $f_\a=Ad(g_{\a\b})f_\b$, with $g_{\a\b}\,:\, U_\a\cap U_\b\rightarrow U(\HS)$ and $Ad(g_{\a\b})\,:\, U_\a\cap U_\b\rightarrow Aut(\KK)=PU(\HS)$. 

We can write the map $\Phi$ as the exponentiation of the C$^*$-algebra of global sections, so that on the open set $U_\a$ it takes the form

\be
\label{emfstk}
\Phi_\a=-e^{i\pi f_\a},
\ee

\noindent where we can take the tachyon field as a representative section of the bundle.

Now we have to impose the $PU(\HS)$-equivariance condition. However, it is not difficult to see that it acts as a gauge transformation that allows us to relate $\Phi_\a$ with the data on $U_\b$ as

\be
\Phi_\a=Ad(g_{\a\b})\Phi_\b,
\ee

In this case, as for the ordinary K-theoretical construction, where the exponential map represented an application from the SUGRA to the Chan-Paton structure, we can consider the exponential in (\ref{emfstk}) as a map from the gerbe structure in the transverse space to the soliton to the group structure in the system of infinite D9-branes given by (\ref{vmwntbf}). In fact, since the only non-vanishing homotopy groups for $U_{cpt}$ are the odd ones, we see again a reflection of all the RR spectrum in type IIA.

In defining the states, we will denote a generic soliton by the pair $({\cal{G}},\Phi)$, where ${\cal{G}}$ is a gauge bundle and $\Phi$ are the automorphism of ${\cal{G}}$. Since the homotopy class of $\Phi$ will characterize the class of these pairs, we will say that this pair is elementary if this class is homotopic to the identity. 

This definition should be handled with care, since the pairs should be considered elementary only in an open set $U_\a$, with a $PU(\HS)$ gauge equivalence relating them. In the context of tachyon condensation, these elementary states correspond to the closed string vacuum. In this case there is a refinement, due to the B-field, that implies that we can establish a second equivalence relation between these different vacua by means of a $PU(\HS)$ transformation, which we can interpret as the effect of a NS5-brane.

This interpretation of D-branes can be traced back to K-homology, by means of the pairing in K-theory
 
\be
{\mbox{\bf{Ext}}}(X,[H])\times K^{-1}(X,[H])\rightarrow \ZZ,
\ee

\noindent given in terms of the index map, where ${\mbox{\bf{Ext}}}(X,[H])$, the twisted group of extensions of $A_{[H]}$ by $\KK$, is the group dual to the K-group\footnote{In the same sense that homology and cohomology are dual theories.}. In particular, every $a\in{\mbox{\bf{Ext}}}(X,[H])$ defines a homomorphism $a_*:K^{-1}(X,[H])\rightarrow \ZZ$. In fact, in \cite{Har.Moo}, it was suggested that the set of type IIA D-branes would be isomorphic to the set of unitary equivalence classes of these extensions.

On the other hand, a description in terms of K-homology, \cite{Asa.Sug.Ter, Sza}, would imply that D-branes are indeed Fredholm modules, \cite{Gra.Var.Fig, Bro, Mat.Sin}, and that the tachyon field is a Fredholm operator (see for example \cite{Boo.Ble}).

This approach is very powerful and would provide a frame to the study of the global anomalies since it not only classifies gauge bundles but also the cycles where D-branes can wrap. However, we leave this study for future research. In any case, in the present context we should be able to see its relevance. In particular, a standard result from string field theory is that in the minima of its potential, the tachyon field should satisfy $T^2=1$.

We can rewrite this condition, by regarding the tachyon field as an operator acting on some state $\phi$, as $(T^2-1)\phi=0$. In K-homology $\phi$ is represented by a set of scalar operators which we can take as parametrizing the codimension space to the D-brane. On the other hand, since we have defined the world-volume in terms of a gauge bundle, we can interpret it as that state satisfying $(T^2-1)\phi\in\KK$, so that we can interpret a D-brane as a compact deformation of the vacuum, defined in terms of the algebra $A_{[H]}$.

\subsubsection{The D6-brane}\label{nontd6}

Now we can apply the previous arguments to our main point of study, the D6-brane. In this case we should compute $K^{-1}(S^3,[H])$, and the charge of this system would be given by the pairing

\be
{\mbox{\bf{Ext}}}(S^3,[H])\times K^{-1}(S^3,[H])\rightarrow \ZZ.
\ee

However, it can be seen, \cite{Mat.Sin}, that taking the DD class as proportional to the volume of the $S^3$ and $N$ an arbitrary integer, ${\mbox{\bf{Ext}}}(S^3,N[H])=0$. This means that for a non-torsion DD class, the system carries no charge.

This result is not surprising, since as in section \ref{globanom} we can interpret this DD class as defining the electric charge of the D6-brane. In this case, due to the geometrical properties of Taub-NUT, and taking the $S^3$ as the Hopf fibration of the eleven dimensional $S^1$ over the $S^2$ at infinity surrounding the core of the monopole, this class is constrained to be torsion. Therefore, the nule charge for a non-torsion DD-class just means that we are not correctly describing a D6-brane.

Moreover, as we will see in section \ref{d6anom}, this result can be interpreted in the context of the cancellation of global anomalies.

\subsection{Anomaly cancellation and $E_8$ bundles}\label{anoe8}

As we have seen, the non-torsion B-field carries gauge data in terms of the group $PU(\HS)$. Therefore, the equivalence class of this field, $[H]\in H^3(X,\ZZ)$, can be seen in terms of the homotopy classes

\be
\label{ddhc}
H^3 (X,\ZZ)=[X,K(\ZZ,3)]=[X,BPU(\HS)].
\ee

\noindent where $K(\ZZ,3)$ is the third Eilenberg-Mac Lane space, which are in general defined for an abelian group $G$ as those spaces satisfying $\pi_m(K(G,n))=G$ for $m=n$ and zero otherwise.

In (\ref{ddhc}), following the DD theorem, we have used the classifying space $BPU(\HS)$ as a model for $K(\ZZ,3)$.  However, it is known that, up to dimension 14, there exists another model for $K(\ZZ,3)$ that has played an important role in string theory, and it is $E_8$, \cite{Wit.6}, so that we can write

\be
\label{ddcaee}
H^3(X,\ZZ)=[X,E_8],
\ee

\noindent from where we can interpret the $E_8$ as the classifying space of the bundles appearing in this section. 

In order to understand the role of $E_8$ in the ten dimensional cancellation of global anomalies, we will define, following \cite{Fre}, the B-field as a differential cocycle (see \cite{Hop.Sin} for background on this topic), i.e. $\check{B}=\check{Z}^3_H(X)$, where

\be
\check{Z}^3_H(X)=\left\{ (c,h,\omega)\in C^3(X,\ZZ)\times C^2(X,\RR)\times \Omega^3(X)\right\},
\ee

\noindent and where $C^*$ and $\Omega^*$ denote as usual cochains and forms respectively.

We can refine this definition and impose that this B-field trivializes the differential cochain associated to the third Stiefel-Whitney class, $W_3\in H^3(X,\ZZ)$, i.e.

\be
d\check{B}=\check{W}_4,
\ee

\noindent where $d\check{B}=(\d c,\omega-c-\d h,d\omega)$, and

\be
\check{W}_4=\left( 0,W_3,0\right)\in C^4(X,\ZZ)\times C^3(X,\RR)\times \Omega^4(X).
\ee

This definition is based on the the interpretation of the eleven dimensional 3-form as a differential cochain, \cite{Dia.Moo.Fre}, and allows us to write the general condition

\be
\label{gacc}
[H]_X-W_3(X)\in H^3(X,\ZZ),
\ee

\noindent which we can relate to the condition for the cancellation of global anomalies (\ref{cnaga}).

The main point is that (\ref{gacc}) can be seen as the dimensionally reduced condition for the cancellation of global membrane anomalies \cite{Wit.7}

\be
[G_4]-W_4\in H^4(Y,\ZZ).
\ee

\noindent where $Y=X\times S^1$ is the eleven dimensional manifold.

Equation (\ref{gacc}) can then be understood as a shifted quantization condition for the B-field, and taken as the DD class in ten dimensions. As we have seen, this class carries the $PU(\HS)\sim K(\ZZ,2)$ gauge information. 

Due to the special topological properties of $E_8$, and since $\pi_n(G)=\pi_{n-1}(\Omega G)$, we can use another model for $K(\ZZ,2)$, up to dimension 14: the based loop group, $\Omega E_8$. Suppose now that in a certain region $Q\subset X$, the DD class vanishes. This leads to

\be
\label{accsdb}
[H]_Q=W_3(Q),
\ee

\noindent which is the condition for the cancellation of global anomalies for a single D-brane wrapping the cycle $Q$, \cite{Fre.Wit} 

An important interpretation of the DD class is as the obstruction to lift a group $G$ to its centrally extended group $\widehat{G}$, i.e. only if $[H]_{DD}=0$ we have a $\widehat{G}$ structure. In our case this implies that (\ref{accsdb}) should be considered as the condition for a  $\widehat{\Omega E}_8$ on $Q$.

At this point we find a subtlety, since from previous considerations we know that in order to interpret (\ref{accsdb}) as the anomaly cancellation condition, the DD class should be torsion. Therefore, we should find a torsion class defining $\widehat{\Omega E}_8$, and, in fact \cite{Pre.Seg}, this class comes from the cocycle that defines the central extension.

In order to compute the homotopy groups associated to ${\widehat{\Omega E}}_8$, we use the long exact homotopy sequence associated to the non-trivial $U(1)$ fibration over $\Omega E_8$. The point in the calculation is that since the cocycle that defines the central extension is an integral 2-form, $\omega\in H^2(\Omega E_8,\ZZ)$, we find that this group is a model for $K(\ZZ_n,1)$, i.e. its only non-trivial homotopy group is $\pi_1(\widehat{\Omega E}_8)=\ZZ_n$. 

From obstruction theory (and as argued in \cite{Ber.Var}), we can associate to this homotopy group a class $y\in H^2(X,\ZZ_n)$, that can be mapped via the Bockstein homomorphism to a class in $H^3(X,\ZZ)$. Therefore, we conclude that the structure of the centrally extended $\widehat{\Omega E}_8$ implies that, in a certain $Q\subset X$, we can have

\bea
\label{caaelg}
[H]_Q-W_3(Q)=0\quad{\mbox{or}}\quad [H]_Q-W_3(Q)=\b(y),
\eea

\noindent where the first case is for $n=1$, corresponding to the basic central extension.

The set of equations (\ref{caaelg}) can be seen as the conditions for the cancellation of global anomalies for the abelian and non-abelian cases. Moreover, from there we can interpret the central charge of $\widehat{\Omega E}_8$ as giving the number of initial unstable D9-branes in the reconstruction of type IIA soliton spectrum, see section \ref{globanom}.

This construction has some other important consequences, the main point being the reduction of the DD class to a torsion class. In the language of gerbes this can be understood as a sort of reduction to a flat B-field (see, for example, \cite{Man} and references therein),i.e. there exists a 1-form that trivializes the B-field in the sense

\bea
\d B_\a & = & d\Lambda_\a,\\
\d A_\a & = & d\rho_{\a\b}+\Lambda_\a,
\eea

\noindent where now the indices label the ${\cal{U}}_\a$ open set covering the manifold.

In standard SUGRA there is no candidate for this $A$ field. However, in Roman's massive SUGRA theory, \cite{Rom}, the 1-form $C^{(1)}$ can play this role since

 \bea
\label{gtft1f}\d B_\a & = & d\Lambda_\a,\\
\label{gtft2f}\d C^{(1)}_\a & = & d\rho_{\a\b}-m\Lambda_\a.
\eea

Of course, the gauge transformation (\ref{gtft2f}) for $C^{(1)}$ implies that it is not a connection due to the $\Lambda$ term. However, we can use the fact that the difference between two different trivializations of a gerbe is a line bundle in order to remove this term. This is a well known mechanism in massive SUGRA theories and is known as the St\"uckelberg mechanism and $C^{(1)}$ is a St\"uckelberg field.

The set of transformations (\ref{gtft1f}) and (\ref{gtft2f}) should be handled with care since one could conclude that $B\propto dC^{(1)}$ which is not the case. In fact, it means that the B-field is flat and that $C^{(1)}$ is a sort of auxiliary field that allows us to define properly a connection on $Q$.

In essence this aims at redefining the B-field as

\be
B\to \tilde B=B+\frac{1}{m}dC^{(1)}
\ee

\noindent so that the new gauge field trivializing $\tilde B$ is indeed a connection. To see this, we write $B=dA$, where $\d A=dh+\Lambda$, and take

\be
\label{gcotcegg}
\tilde A=A+\frac{1}{m}C^{(1)}.
\ee

With this definition, the new 2-form $\tilde B$ is gauge invariant and $\tilde A$ is a connection since it can be shown that its gauge transformation is

\be
\d\tilde A=dh+\frac{1}{m}d\rho=-id\log k,
\ee

\noindent where we have defined $h=-i\log f$ and $\rho=-i\log g$ so that
 
\be
\label{tffcegg}
k_{\a\b}=f_{\a\b}\cdot g_{\a\b}^{1/m}.
\ee

From the previous interpretation of the central charge of $\widehat{\Omega E}_8$ as giving the number of D9-branes and, in the spirit of \cite{Kap}, we can interpret (\ref{gcotcegg}) as the gauge connection of the $\widehat{\Omega E}_8$ bundle and (\ref{tffcegg}) as its transition functions.

From here we see that $C^{(1)}$ is indeed the connection on the line bundle that defines the central extension and the central charge is therefore Romans' mass parameter. In this sense, the mass parameter gives us the rank of the gauge theory underlying the K-theoretical classification of string solitons.

In the previous discussion, we have only cared about the effects of the B-field and the based loop group. However, in the dimensional reduction we have the free loop group, $LE_8$ centrallyetrally extended group $\widehat{LE}_8$ (see \cite{Gar.Mur, Evs.1, Mat.Sat, Ber.Var} for different discussions concerning this point). In any case, it is clear that the cancellation of global anomalies leads to the centrally extended group, defined as the semidirect product $\widehat{LE}_8=E_8\ltimes\widehat{\Omega E}_8$, which in this case means that $\widehat{LE}_8$ is the trivial fibration of $\widehat{\Omega E}_8$ over $E_8$.

This suggests that, as already pointed in \cite{Man}, we should modify\footnote{It should be stressed again that we are only concerned about topology.} the gauge group in (\ref{vmwntbf}) as

\be
\label{ggce}
{\widehat{\cal{G}}}=\frac{U_{cpt}}{\widehat{LE}_8},
\ee

\subsubsection{The D6-brane}\label{d6anom}

Using this $E_8$ formalism we can see the spectrum of solitons in usual type IIA and IIB, \cite{Ada.Evs.1, Evs.1}, as well as in massive SUGRA \cite{Evs.1, Ber.Var}. Now we will relate it to twisted K-theory, in the particular case of the D6-brane.

The interpretation carried in section \ref{globanom} concerning this soliton and eleven dimensions implied that the K-theoretical $\pi_3$ is a reflection of the eleventh dimension. In the previous context, we could associate it to the $\pi_3(E_8)$, where the 3-sphere comes from the Hopf fibration of Taub-NUT space. 

However, this picture is consistent when we include a non-torsion B-field, and as we have mentioned in section \ref{nontd6}, the index pairing in twisted K-theory gives no charge for this system. Therefore we need further restrictions, which come from the previous analysis for the cancellation of global anomalies.

In the presence of a D6-brane, we can consider the decomposition (\ref{C}) and write the dimensional reduction of the eleven dimensional 4-form as

\be
G^{(4)}=dC^{(3)}=G_{ijkl}+G_{\m\n\r\s}+H_{ijk}V_\m+\tilde H_{\m\n\r}A_i+B_{ij}dV_\m+dA_iC_{\m\n},
\ee

\noindent where $H_{ijk}=dB_{ij}$ and $\tilde H_{\m\n\r}=dC_{\m\n}$. If we take, as in \cite{Spa} a 4-cycle, $V$, along the directions $(i,j,\m,x^4)$ (which can be interpreted as a membrane wrapping the Taub-NUT direction) and integrate $G^{(4)}$ along it, we obtain

\be
\label{hfmas}
\int_V G^{(4)}=c\int_U\left(B+dA\right),
\ee

\noindent where $U$ denotes the 2-cycle in the D6-brane world-volume and $c$ is just the numerical factor that arises in the integration of $C_{\m\n}$.

Of course, for the D6-brane (\ref{hfmas}) vanishes, since the KK-monopole in eleven dimensions is a purely gravitational solution with $G^{(4)}=0$, and this translates in the torsion property of the B-field in its world volume in terms of the A-field, \cite{Gom.Man.1}. Anyway, (\ref{hfmas}), is valid in general, and it has a very nice interpretation since implies that the change in the holonomy of the eleven dimensional 3-form around the membrane world volume is the same that arises in the string partition function.

An interesting fact is that we could write, formally, this structure as a reduction from $H^2(LM,\ZZ)$ to $tor\left(H^2(LQ,\ZZ)\right)$, where $Q$ denotes the D6-brane world volume.  In this sense, the tachyon condensation gives a map from the $U_{cpt}/{\widehat{LE}_8}$ structure in the world volume of the system of D9-branes, into that of a flat gerbe in the world volume (and in the transverse space).

\section{Conclusions and comments}

In this paper we have studied the relation between ten and eleven dimensional physics as given in K-theory. 

Our starting point has been the possible existence of an electric charge for the D6-brane, as predicted in K-theory. Indeed we have found that the electric degrees of freedom can be associated to the C-monopole, i.e. to the state carrying the eleven dimensional 3-form, $C^{(3)}$, charge, in a process similar to the ten dimensional KK-monopole, which gets an electric charge from the H-monopole, being the only difference that now we obtain information about the world-volume theory.

One of the main points in the paper has been the proper reconstruction of the moduli space of this gravitational monopole, as an $SU(2)$ monopole, which, despite the inclusion of the B-field, does not induce any non-commutative deformation. This absence of deformation is realized in terms of the vanishing of certain characteristic class associated to the 't Hooft flux and determines a Spin$^c$ structure.

From this Spin$^c$ structure and the conditions for the cancellation of the Freed-Witten anomaly, we have constructed an equivalence map between the $U(2)$ isometry group of rotations in the transverse space to the KK-monopole in eleven dimensions and the $U(2)$ Chan-Paton gauge group in the world-volume of the two unstable D9-branes needed in the K-theoretical construction.

However, since our starting point is a system of unstable D9-branes, we should not consider any restriction on the properties of the B-field. In particular, this means that we should consider non-torsion B-fields. In doing so, we are led to consider a system of infinite D9-branes whose world-volume theory has the structure of a gauge bundle with $U_{cpt}/PU(\HS)$ gauge group.

In any case, whatever initial conditions we chose, the final state after the tachyon condensation takes place, should be the same state or, at least, a state that shares the main properties. In particular, for the D6-brane we require a codimension four (in eleven dimensions) anomaly free soliton with torsion B-field.

As we have seen, these conditions will still lead to a connection between the ten and the eleven dimensional data, but in this case, the equivalent to the Spin$^c$ structure is that of an $E_8$ gauge theory. Therefore, we see that including the B-field modifies the general structure of the theory in a very non-trivial way. 

Once again a main tool in our study has been the cancellation of global anomalies in ten dimensions which we have seen to lift directly to the condition for the membrane anomaly and allowed us to reinterpret these conditions as a shifted quantization condtion for the B-field and the field itself as a shifted differential character, in analogy with \cite{Dia.Moo.Fre}, where this was the condition for the eleven dimensional 3-form.

In this case, however, there is a subtlety concerning the large-N limit. Namely, while for the standard and torsion twisted K-theory classifications it seems that we only need a finite number of D9-branes, in the non-torsion case we are strictly in the $N=\infty$ limit. Therefore, in the process of tachyon condensation we should recover the whole spectrum of type IIA, and also a torsion B-field.
 
Once we implement this property in the shifted quantization condition for the B-field we have found that the torsion is directly related to the Romans' mass and that this mass gives precisely the number of D9-branes of the initial state, i.e. is the rank of the $U(n)$ gauge theory on the world-volume of the D9-branes.

There are, of course, some open problems. First, as we mentioned in the introduction, a proper understanding of the K-homology role in the anomaly cancellation. On the other hand, following \cite{Egu.Gil.Han}, the KK-monopole should be considered as a manifold with two boundaries rather than as a closed manifold. It could be interesting to find out the proper role of these boundaries in this $E_8$ formalism.

\section*{Acknowledgments}

I am indebted to C.~G\'omez for many useful conversations and for a careful reading of a version of this paper. I would like to thank also N.~Alonso-Alberca, J.~Evslin, J.~Gonzalo and M.~Logares for many helpful conversations. I would like to thank also N.~Hitchin, J.~Louis and T.~Ort\'\i n for useful comments.

\begin{appendix}

\section{The isomorphism $\left[I,\a\right]\sim K({\cal{S}}'(X))$}\label{k1tok}

As we have mentioned (see for example \cite{Hor}, \cite{Ols.Sza} and \cite{Man}), there are two different, but completely equivalent, definitions of the $K^{-1}(X)$ group. These definitions are commonly known in the physics literature as the M-theoretical and the stringy definitions, since the first one needs an extension of the manifold while the second is defined in terms of the elements defined in the manifold with no need of extending it. The purpose of this appendix is to review the relation between these two definitions, see \cite{Ati, Kar} for a rigorous demonstration.

Following \cite{Ati}, we define the higher K-group $K^{-1}(X)$ in terms of the suspension ${\cal{S}}'(X)$ as

\be
K^{-1}(X)=K({\cal{S}}'(X)).
\ee

\noindent This is the "M-theoretical" definition.

However, in the body of the paper we have used the "stringy" definition, in which case, the equivalence classes are determined from pairs $(I,\a)$, where $I$ are trivial vector bundles on a manifold $X$ and $\a$ is an automorphism of $I$, in terms of the homotopy classes of $\a$. In this way, we denote a class as trivial if $\a$ is homotopic to the identity.

The key point in the identification of both definitions is the fact that we can see the automorphism $\a$ as defining the transition functions of a vector bundle, $E_\a$, over ${\cal{S}}'(X)$, whose isomorphism classes are classified by $K({\cal{S}}'(X))$, while their are also in one-to-one correspondence with the homotopy class of $\a$.

Although we will deal with the simpler case when the manifold $X$ is an $S^n$, let us begin by defining on general grounsuspensionpension ${\cal{S}}'(X)$. In doing this we need the concept of smash product, $X\wedge Y$, of topological spaces $X$ and $Y$

\be
X\wedge Y= X\times Y/(X\vee Y).
\ee

\noindent This means that $X\wedge Y$ is obtained from $X\times Y$ by shrunking $X\vee Y=\left( x_0\times Y\right)\cup \left( X\times y_0\right)$ to a point, where $x_0$ and $y_0$ are the base points of $X$ and $Y$ respectively. Then, the n-th iterated reduced suspension of $X$ is defined as

\be
{\cal{S}}'(X)=S^1\wedge S^1\dots\wedge S^1\wedge X=S^n\wedge X,
\ee

\noindent where we have used a very important fact for the following, and it is that there is a natural homeomorphism

\be
S^n\approx S^1\wedge S^1\dots\wedge S^1,
\ee

\noindent with $n$ factors.

Now we can define the second abstract concept and it is the double cone of $X$. It is defined as the quotient of $X\times [-1,1]$ by the equivalence relation which identifies $X\times \left\{ 1\right\}$ with a single point and $X\times \left\{ -1\right\}$ with another single point. 

Then, denoting by $C_+X$ the image of $X\times [0,1]$ (and by $C_-X$ the image of $X\times [-1,0]$) in the quotient, we can write

\be
{\cal{S}}'(X)=C_+X\cup C_-X,
\ee

\noindent where, obviously

\be
X=C_+X\cap C_-X.
\ee

From these expressions it is clear that an automorphim of a vector bundle over $X$ will be seen as a transition function of this vector bundle over ${\cal{S}}'(X)$. Let us see it in the case of $S^n$.

Let us take the unit sphere $S^n$ of $\RR^n$, $S^n=\left\{x\in\RR^{n+1}|\;||x||\leq 1\right\}$, and consider its two hemispheres

\bea
S^n_+=\left\{x\in S^n|\; x_{n+1}\geq 1\right\},\qquad S^n_-=\left\{x\in S^n|\; x_{n+1}\leq 1\right\}
\eea

Since $S^n=S^1\wedge S^{n-1}$, its is not difficult to see that these two sets correspond to the $C_+X$ and $C_-X$, and $S^{n-1}=S^n+\cap S^n_-$ corresponds to our manifold $X$. 

Suposse now, that we are given a function, $f$, on the intersection, $S^{n-1}$

\be
\label{mf}
f\,:\, S^{n-1}\longrightarrow GL_p(k),
\ee

\noindent where $k=(\RR, \CC,...)$ and $p$ is the rank. Then, we can define a bundle $E_f$ over ${\cal{S}}'(X)=S^n$ in terms of the following clutching of trivial bundles. Let $E_1$ and $E_2$ be two trivial vector bundles over $S^n_+$ and $S^n_-$ respectively, i.e.

\bea
E_1 & = & S^n_+\times k^p, \\
E_2 & = & S^n_-\times k^p.
\eea

\noindent these bundles are clutched by the transition function

\be
g_{21}=\hat f\,:\, S^{n-1}\times k^p\longrightarrow S^{n-1}\times k^p,
\ee

\noindent where $\hat f$ is the map that induces $f$, given by (\ref{mf}), on each fiber.

The bundle $E_f$ is then defined as the vector bundle over $S^n$ with local trivializations given by

\be
h_\pm\,:\, \left. E_f\right|_{S^n_{\pm}}\longrightarrow S^n_{\pm}\times k^p,
\ee

\noindent so that $g_{21}=\hat f=h_+\circ h_-^{-1}$ on $S^{n-1}$, satisfying the ususal cocycle relation.

Now, as usual, the equivalence classes of isomorphisms of this trivial vector bundle over $S^n$ are classified by $K(S^n)$, but what is more important for us is that these equivalence classes descend to the equivalence classes of isomorphism of the vector bundle defined by $f$ over $S^{n-1}$, and these only depend on the homotopy class of $f$ (see, for example, lemma 1.4.7 in \cite{Ati}), establishing the desired correspondence.

\end{appendix}


\begin{thebibliography}{10}

\bibitem{Min.Moo}
R.~Minasian and G.W.~Moore.
\newblock K-theory and Ramond-Ramond charge.
\newblock {\em JHEP}, 11:002, 1997.

\bibitem{Wit.1}
E.~Witten.
\newblock D-branes and K-theory.
\newblock {\em JHEP}, 12:019, 1998.

\bibitem{Hor}
P.~Horava.
\newblock Type IIA D-branes, K-theory, and Matrix theory.
\newblock {\em Adv. Theor. Math. Phys.}, 2:1373--1404, 1999.

\bibitem{Moo.Wit}
G.W.~Moore and E.~Witten.
\newblock Self-duality, Ramond-Ramond fields, and K-theory.
\newblock {\em JHEP}, 05:032, 2000.

\bibitem{Fre.Hop}
D.S.~Freed and M.J.~Hopkins.
\newblock On Ramond-Ramond fields and K-theory.
\newblock {\em JHEP}, 05:044, 2000.

\bibitem{Fre.Wit}
D.S.~Freed and E.~Witten.
\newblock Anomalies in string theory with D-branes.
\newblock hep-th/9907189.

\bibitem{Mal.Moo.Sei.1}
J.~Maldacena, G.W.~Moore, and N.~Seiberg.
\newblock D-brane instantons and K-theory charges.
\newblock {\em JHEP}, 11:062, 2001.

\bibitem{Sen.1}
A.~Sen.
\newblock Stable non-BPS states in string theory.
\newblock {\em JHEP}, 6:007, 1998.

\bibitem{Sen.2}
A.~Sen.
\newblock Stable non-BPS bound states of BPS-branes.
\newblock {\em JHEP}, 8:010, 1998.

\bibitem{Sen.4}
A.~Sen.
\newblock Tachyon condensation on the brane antibrane system.
\newblock {\em JHEP}, 08:012, 1998.

\bibitem{Sen.5}
A.~Sen.
\newblock Universality of the tachyon potential.
\newblock {\em JHEP}, 12:027, 1999.

\bibitem{Ols.Sza}
K.~Olsen and R.J.~Szabo.
\newblock Constructing D-branes from K-theory.
\newblock {\em Adv. Theor. Math. Phys.}, 3:889--1025, 1999.

\bibitem{Man}
J.J.~Manjar\'\i n.
\newblock Topics on D-brane charges with B-fields.
\newblock {\em Int. J. Geom. Meth. Mod. Phys.}, 1:N4, 2004.

\bibitem{Wit.4}
E.~Witten.
\newblock Overview of K-theory applied to strings.
\newblock {\em Int. J. Mod. Phys.}, A16:693--706, 2001.

\bibitem{Wit.5}
E.~Witten.
\newblock Noncommutative tachyons and string field theory.
\newblock hep-th/0006071.

\bibitem{Tow.1}
P.K.~Townsend.
\newblock The eleven-dimensional supermembrane revisited.
\newblock {\em Phys. Lett.}, B350:184--187, 1995.

\bibitem{Tow.2}
P.K.~Townsend.
\newblock M-theory from its superalgebra.
\newblock hep-th/9712004.

\bibitem{Ber.Jan.Ort}
E.~Bergshoeff, B.~Janssen, and T.~Ort\'\i n.
\newblock Kaluza-klein monopoles and gauged sigma-models.
\newblock {\em Phys. Lett.}, B410:131--141, 1997.

\bibitem{Ima}
Y.~Imamura.
\newblock Born-infeld action and Chern-Simons term from Kaluza-Klein monopole
  in M-theory.
\newblock {\em Phys. Lett.}, B414:242--250, 1997.

\bibitem{Gom.Man.2}
C.~G\'omez and J.J.~Manjar\'\i n.
\newblock Dyons, K-theory and M-theory.
\newblock hep-th/0111169.

\bibitem{Gom.Man.1}
C.~G\'omez and J.J.~Manjar\'\i n.
\newblock A note on the dyonic D6-brane.
\newblock {\em Int. J. Mod. Phys.}, A19, Supplementary Issue 2:217, 2004.

\bibitem{Jul.Zee}
B.~Julia and A.~Zee.
\newblock Poles with both magnetic and electric charges in nonabelian gauge
  theory.
\newblock {\em Phys. Rev.}, D11:2227--2232, 1975.

\bibitem{Nah.1}
W.~Nahm.
\newblock The construction of all selfdual multi-monopoles by the ADHM
  method. (talk).
\newblock in Monopoles in Quantum Field Theory, Proceedings of the monopole meeting in Trieste Cent. Theor. Phys. - IC-82-016 (82,REC.MAR.) 8p, World-Scientific, Singapore, 1982.

\bibitem{Nah.2}
W.~Nahm.
\newblock All selfdual multi-monopoles for arbitrary gauge groups.
\newblock Presented at Int. Summer Inst. on Theoretical Physics, Freiburg, West
  Germany, Aug 31 - Sep 11, 1981, (Preprint) TH 3172-CERN.

\bibitem{Nah.3}
W.~Nahm.
\newblock The algebraic geometry of multi-monopoles.
\newblock Presented at 11th Int. Colloq. on Group Theoretical Methods in
  Physics, Istanbul, Turkey, Aug 23-28, 1982, (Preprint) Physikalisches Institut, University of Bonn.

\bibitem{Hit.2}
N.J.~Hitchin.
\newblock On the construction of monopoles.
\newblock {\em Commun. Math. Phys.}, 89:145--190, 1983.

\bibitem{Don}
S.K.~Donaldson.
\newblock Nahm's equations and the classification of monopoles.
\newblock {\em Commun. Math. Phys.}, 96:387--407, 1984.

\bibitem{Don.Kar}
P.~Donovan and M.~Karoubi.
\newblock Graded Brauer groups and K-theory with local coefficients.
\newblock {\em IHES Pub.}, 38:4, 1970.

\bibitem{Kap}
A.~Kapustin.
\newblock D-branes in a topologically nontrivial B-field.
\newblock {\em Adv. Theor. Math. Phys.}, 4:127--154, 2000.

\bibitem{Ros}
J.~Rosenberg.
\newblock Continuous-trace algebras from the bundle theoretical point of view.
\newblock {\em J. Austral. Math. Soc.}, Series A 47:368, 1989.

\bibitem{Bou.Mat}
P.~Bouwknegt and V.~Mathai.
\newblock D-branes, B-fields and twisted K-theory.
\newblock {\em JHEP}, 03:007, 2000.

\bibitem{Ada.Evs.1}
A.~Adams and J.~Evslin.
\newblock The loop group of $E_8$ and K-theory from 11d.
\newblock {\em JHEP}, 02:029, 2003.

\bibitem{Evs.1}
J.~Evslin.
\newblock From $E_8$ to F via T.
\newblock {\em JHEP}, 08:021, 2004.

\bibitem{Mat.Sat}
V.~Mathai and H.~Sati.
\newblock Some relations between twisted K-theory and $E_8$ gauge theory.
\newblock {\em JHEP}, 03:016, 2004.

\bibitem{Gar.Mur}
H.~Garland and M.K.~Murray.
\newblock Kac-moody monopoles and periodic instantons.
\newblock {\em Commun. Math. Phys.}, 120:335--351, 1988.

\bibitem{Ber.Var}
A.~Bergman and U.~Varadarajan.
\newblock Loop groups, Kaluza-Klein reduction and M-theory.
\newblock hep-th/0406218.

\bibitem{Bro}
J.~Brodzki.
\newblock An introduction to K-theory and cyclic cohomology.
\newblock funct-an/9606001.

\bibitem{Har.Moo}
J.A.~Harvey and G.W.~Moore.
\newblock Noncommutative tachyons and K-theory.
\newblock {\em J. Math. Phys.}, 42:2765--2780, 2001.

\bibitem{Mat.Sin}
V.~Mathai and I.M.~Singer.
\newblock Twisted K-homology theory, twisted ext-theory.
\newblock hep-th/0012046.

\bibitem{Sza}
R.J.~Szabo.
\newblock D-branes, tachyons and K-homology.
\newblock {\em Mod. Phys. Lett.}, A17:2297--2316, 2002.

\bibitem{Sen.3}
A.~Sen.
\newblock Kaluza-klein dyons in string theory.
\newblock {\em Phys. Rev. Lett.}, 79:1619--1621, 1997.

\bibitem{Gre.Har.Moo}
R.~Gregory, J.A.~Harvey, and G.W.~Moore.
\newblock Unwinding strings and T-duality of Kaluza-Klein and H-monopoles.
\newblock {\em Adv. Theor. Math. Phys.}, 1:283--297, 1997.

\bibitem{Hul}
C.M.~Hull.
\newblock Gravitational duality, branes and charges.
\newblock {\em Nucl. Phys. Proc. Suppl.}, 62:412--421, 1998.

\bibitem{Ati.Hit}
M.F.~Atiyah and N.J.~Hitchin.
\newblock {\em The geometry and dynamics of magnetic monopoles}.
\newblock Princeton Univ. Press, New York, 1988.

\bibitem{Hit}
N.~Hitchin.
\newblock Lectures on special lagrangian submanifolds.
\newblock math.DG/9907034.

\bibitem{Dub.Fom.Nov}
B.A.~Dubrovin, A.T.~Fomenko, and S.P.~Novikov.
\newblock {\em Modern geometry methods and applications, pt. III}.
\newblock Springer, 1990.

\bibitem{Ati.Bot.Sha}
M.~Atiyah, R.~Bott, and A.~Shapiro.
\newblock Clifford modules.
\newblock {\em Topology} 1:245, 1962.

\bibitem{Dia.Moo.Wit.1}
D.~Diaconescu, G.W.~Moore, and E.~Witten.
\newblock $E_8$ gauge theory, and a derivation of K-theory from M-theory.
\newblock {\em Adv. Theor. Math. Phys.}, 6:1031--1134, 2003.

\bibitem{Dix.Dou}
J.~Dixmier and A.~Douady.
\newblock Champs continus d'spaces hilbertiens et de c$^*$-alg\`ebres.
\newblock {\em Bull. Soc. Math. France}, 91:227, 1963.

\bibitem{Har}
J.A.~Harvey.
\newblock Topology of the gauge group in noncommutative gauge theory.
\newblock hep-th/0105242.

\bibitem{Car.Mic}
A.L.~Carey and J.~Mickelsson.
\newblock The universal gerbe, dixmier-douady class, and gauge theory.
\newblock {\em Lett. Math. Phys.}, 59:47--60, 2002.

\bibitem{Pal}
R.S.~Palais.
\newblock On the homotopy type of certain groups of operators.
\newblock {\em Topology}, 3:271, 1965.

\bibitem{Gop.Min.Str}
R.~Gopakumar, S.~Minwalla, and A.~Strominger.
\newblock Noncommutative solitons.
\newblock {\em JHEP}, 05:020, 2000.

\bibitem{Das.Muk.Raj}
K.~Dasgupta, S.~Mukhi, and G.~Rajesh.
\newblock Noncommutative tachyons.
\newblock {\em JHEP}, 06:022, 2000.

\bibitem{Kar}
M.~Karoubi.
\newblock {\em K-theory}.
\newblock Springer-Verlag, 1978.

\bibitem{Asa.Sug.Ter}
T.~Asakawa, S.~Sugimoto, and S.~Terashima.
\newblock D-branes, Matrix theory and K-homology.
\newblock {\em JHEP}, 03:034, 2002.

\bibitem{Gra.Var.Fig}
J.C.~V\'arilly J.M.~Gracia-Bond{\'\i}a and H.~Figueroa.
\newblock {\em Elements of noncommutative geometry}.
\newblock Birkh\"auser Advanced Texts, 2000.

\bibitem{Boo.Ble}
B.~Booss and D.D.~Bleecker.
\newblock {\em Topology and Analysis: The Atiyah-Singer Index Formula and
  Gauge-Theoretic Physics}.
\newblock Springer-Verlag, 1977.

\bibitem{Wit.6}
E.~Witten.
\newblock Topological tools in ten-dimensional physics.
\newblock {\em Int. J. Mod. Phys.}, A1:39, 1986.

\bibitem{Fre}
D.S.~Freed.
\newblock Dirac charge quantization and generalized differential cohomology.
\newblock hep-th/0011220.

\bibitem{Hop.Sin}
M.J.~Hopkins and I.M.~Singer.
\newblock Quadratic functions in geometry, topology, and M-theory.
\newblock math.AT/0211216.

\bibitem{Dia.Moo.Fre}
D.~Diaconescu, G.W.~Moore, and D.S.~Freed.
\newblock The M-theory 3-form and $E_8$ gauge theory.
\newblock hep-th/0312069.

\bibitem{Wit.7}
E.~Witten.
\newblock On flux quantization in M-theory and the effective action.
\newblock {\em J. Geom. Phys.}, 22:1--13, 1997.

\bibitem{Pre.Seg}
A.~Pressley and G.~Segal.
\newblock {\em Loop groups}.
\newblock Oxford Mathematical Publications, New York, 1986.

\bibitem{Rom}
L.J.~Romans.
\newblock Massive N=2A supergravity in ten-dimensions.
\newblock {\em Phys. Lett.}, B169:374, 1986.

\bibitem{Spa}
J.~Sparks.
\newblock Global worldsheet anomalies from M-theory.
\newblock {\em JHEP}, 08:037, 2004.

\bibitem{Egu.Gil.Han}
T.~Eguchi, P.B.~Gilkey, and A.J.~Hanson.
\newblock Is the Taub-NUT metric a gravitational instanton?
\newblock {\em Phys. Rev.}, D17:423--427, 1978.

\bibitem{Ati}
M.~Atiyah.
\newblock {\em K-theory}.
\newblock W.A. Benjamin, New York, 1967.

\end{thebibliography}
\end{document}